\documentclass[english,3p]{elsarticle}
\usepackage[T1]{fontenc}
\usepackage[latin9]{inputenc}
\usepackage{amsmath}
\usepackage{amssymb}
\usepackage{amstext}
\usepackage{float}
\usepackage{graphicx}
\usepackage{babel}

\begin{document}

\title{Controlling multistability in coupled systems with soft impacts}

\author[rvt]{P. Brzeski}

\author[rvt1]{E. Pavlovskaia}

\author[rvt]{T. Kapitaniak}

\author[rvt]{P. Perlikowski\corref{cor1}}

\ead{przemyslaw.perlikowski@p.lodz.pl}

\cortext[cor1]{Corresponding author}

\address[rvt]{Division of Dynamics, Lodz University of Technology, Stefanowskiego 1/15, 90-924 Lodz, Poland}

\address[rvt1]{Centre for Applied Dynamics Research, School
of Engineering, University of Aberdeen, AB24 3UE, Aberdeen, Scotland}

\begin{abstract}
In this paper we present an influence of discontinuous coupling on
the dynamics of multistable systems. Our model consists of two periodically
forced oscillators that can interact via soft impacts. The controlling
parameters are the distance between the oscillators and the difference
in the phase of the harmonic excitation. When the distance is large
two systems do not collide and a number of different possible solutions
can be observed in both of them. When decreasing of the distance,
one can observe some transient impacts and then the systems settle
down on non-impacting attractor. It is shown that with the properly chosen
distance and difference in the phase of the harmonic excitation, the number of possible solutions of the coupled systems can be reduced.
The proposed method is robust and applicable in many different systems.
\end{abstract}


\maketitle

\section{Introduction}

The interaction between impacting systems is nowadays extensively
investigated. In many systems such as tooling machines, gear boxes, heat
exchangers and backlash gear the motion of some elements is limited
by a barrier. There are many impact models which give the relations
between the interacting system elements. Generally, they can be divided
into two groups, i.e., the hard and soft impacts \cite{moreau1988nonsmooth,gilardi2002literature,Hutzler2004,brogliato1999nonsmooth}.
The hard impacts are modeled by the restitution coefficient \cite{kundu2011vanishing,witelski2014driven,blazejczyk2010hard}.
In this approach the time of contact is infinitely small and the exchange
of energy is instantaneous. The second approach (soft impact) assumes
the finite, nonzero contact time and a penetration of the base by
the colliding body. Hence, the contact is modeled as a linear \cite{shaw1983periodically,andreaus2010numerical,zhang2011multi},
Hertzian \cite{goldsmith1964impact,serweta2014lyapunov} or other
nonlinear \cite{peterka2003behaviour} spring and viscous damper.
The separate equations of motion describe the in-contact and out-of-contact
dynamics.

The numerous works have been devoted to phenomena induced by the the impacts. The bifurcation scenarios and implication of grazing
events are quite well understood \cite{Banerjee1996,bernardo2008piecewise,ganguli2005dangerous,ma2006border}.
There are a few studies which focus on the systems where impacts between
coupled oscillators are transient. Blazejczyk-Okolewska \textit{et.
al.} \cite{blazejczyk2001antiphase} show that impacting systems can
synchronize (via the exchange of energy during the contact) in anti-phase
on chaotic attractor. The impacts can be considered as a discontinuous
transient coupling which disappears once the interacting systems reach
the synchronous solutions.

The phenomenon of synchronization is commonly encountered in non-linear systems
\cite{Pikovsky2001,Strogatz2005,Perlikowski2008a}. Generally, in coupled mechanical systems one can
observe two types of synchronous motion, i.e., the complete and the
phase synchronization \cite{Blekhman1988,Fujisaka1983,pikovsky1984,Kapitaniak20141}. As the coupling between mechanical oscillators (two directly interacting bodies or via spring, damper or inerter) is always bidirectional; when systems are completely synchronized the
value of coupling force is equal to zero, and only if common motion
is perturbed the systems once again start to interact (note that for non-mechanical systems it is not always true). This is the straightforward
analogy to the above mentioned discontinuous transient coupling via
impacts, where the perturbation of stable non-impacting solution leads
to the appearance of transient impacting motion (coupling).

In this paper we demonstrate the idea and present solution to reduce complexity via
transient impacts. We consider systems of two identical oscillators and assume
that interaction between them occurs through soft impacts. When
the systems are uncoupled we observe multiple stable attractors for
each subsystem. Using a piecewise transient coupling to another
identical subsystem we can change the number of stable attractors
and, in many cases, specify on which attractor both systems settle.

The paper is organized as follows. In Section 2 we consider a simple
model which is used to demonstrate the main idea of our approach and
define the notations introduced to describe existing periodic states.
In the next section we present and describe the how via discontinuous
coupling we can decrease number of solutions in the complex systems
with many co-existing periodic solutions of different type. The possible
coexistence of impacting and non-impacting solutions is discussed
in Section 4. Finally, in Section 5 the conclusions are given.

\section{Duffing systems}

In this section we show the main idea of our approach using a simple
example. First, the considered model is introduced and the notations
used to classify the existing solutions are defined. Then, we present
the results of the numerical analysis.

\subsection{Model of the system}

\begin{figure}[H]
\centering{}
\includegraphics{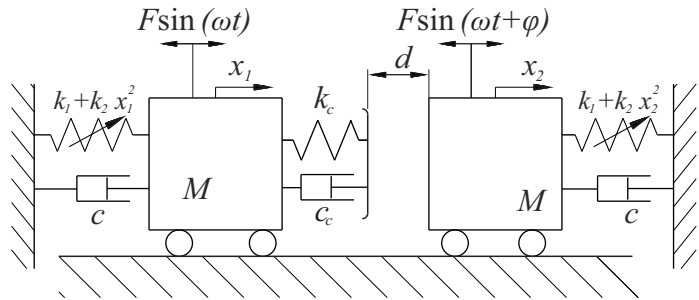}
\protect\protect\protect\caption{Model of two discontinuously coupled Duffing oscillators. \label{fig:Figure1}}
\end{figure}

The system considered in this subsection consists of two identical
Duffing oscillators shown in Figure \ref{fig:Figure1}. Oscillators in their steady states (at rest)
are separated by the distance $d$ , and the impacts which could occur
between them are of the soft type due to the presence of a spring
with stiffness $k_{c}$ and a viscous damper with damping coefficient
$c_{c}$. Two Duffing oscillators have masses $M$ each and are damped
by viscous dampers with damping coefficient $c$. The spring connecting
each oscillator to the wall is nonlinear and of hardening type, where
both stiffness coefficients are positive: $k_{1}>0$ and $k_{2}>0$.
Both Duffing systems are driven by harmonic forces with the amplitude
$F$ and the frequency $\omega$ but there is a phase shift between
these forces. Forcing of the first oscillator has fixed phase (equal
to zero) while for the second one there is a phase shift $\varphi$
which is used as a control parameter ($\varphi\in\left\langle 0,\:2\pi\right)$).

The whole system is described by the following equations of motion:

\begin{equation}
M\ddot{x}_{1}+k_{1}x_{1}+k_{2}x_{1}^{3}+c\dot{x}_{1}+F_{C}=F\sin\left(\omega t\right)\label{eq:Duff_1}
\end{equation}

\begin{equation}
M\ddot{x}_{2}+k_{1}x_{2}+k_{2}x_{2}^{3}+c\dot{x}_{2}-F_{C}=F\sin\left(\omega t+\varphi\right)\label{eq:Duff_1-1}
\end{equation}
where $F_{C}$ describes the forces generated by the discontinuous
coupling and is given by the formula:
\begin{equation}
F_{C}=\begin{cases}
\begin{array}{cc}
0\\
\\
k_{c}\left(\left(x_{1}-x_{2}\right)-d\right)+c_{c}\left(\dot{x}_{1}-\dot{x}_{2}\right)
\end{array} & \begin{array}{cc}
for & x_{1}-x_{2}<d\\
\\
for & x_{1}-x_{2}\geqslant d
\end{array}\end{cases}\label{eq:Duff_1-1-1}
\end{equation}
The values of the parameters are as follows: $M=1.0\,[kg]$, $k_{1}=1.0\,[\frac{N}{m}]$,
$k_{2}=0.01\,[\frac{N}{m^{3}}]$, $c=0.05\,[\frac{Ns}{m}]$, $F=1.0\,[N]$,
$\omega=1.3\,[\frac{1}{s}]$, $k_{c}=8.0\,[\frac{N}{m}]$, $c_{c}=10.0\,[\frac{N}{m}]$. Distance between system $d$ and phase shift in excitation $\varphi$ are controlling parameters. Introducing dimensionless time $\tau=t\omega_{1}$,
where $\omega_{1}=1\,[\frac{1}{s}]$, reference length $l_{r}=1.0\,[m]$
and mass $m_{r}=1\,[kg]$ we transform the equations (\ref{eq:Duff_1}) -- (\ref{eq:Duff_1-1-1}) into dimensionless form in which dimensional parameters are replaced by the
following non-dimensional parameters:

\begin{center}
$M^{\prime}=\frac{M}{m_{r}}$,
$k_{1}^{\prime}=\frac{k_{1}l_{r}}{m_{r}\omega_{1}^{2}}$, $k_{2}^{\prime}=\frac{k_{2}l_{r}^{2}}{m_{r}\omega_{1}^{2}}$,
$c^{\prime}=\frac{c}{m_{r}\omega_{1}}$, $F^{\prime}=\frac{F}{m_{r}l_{r}\omega_{1}^{2}}$,
$\omega^{\prime}=\frac{\omega}{\omega_{1}}$, $k_{c}^{\prime}=\frac{k_{c}l_{r}}{m_{r}\omega_{1}^{2}}$,
$c_{c}^{\prime}=\frac{c_{c}}{m_{r}\omega_{1}}$, $d^{\prime}=\frac{d}{l_{r}}$.
\end{center}
We perform transformation to the dimensionless form in the way
that enables to hold the values of parameters, hence: $M^{\prime}=1.0$,
$k_{1}^{\prime}=1.0$, $k_{2}^{\prime}=0.01$, $c^{\prime}=0.05$,
$F^{\prime}=1.0$, $\omega^{\prime}=1.3$, $k_{c}^{\prime}=8.0$,
$c_{c}^{\prime}=10.0$. For simplicity all of
the primes used in definitions of dimensionless parameters will be
omitted hereafter in the analysis.

\subsection{Notations for the periodic solutions}

We introduce the notations that enable to describe all periodic
states of two impacting oscillators, hence we can classify all solutions
that can occur in the considered system. To recall, we assume that
the left (first) system is a reference system, hence its phase of
excitation and position are fixed while for the right (second) system
the phase of excitation $\varphi$ can vary in the range form $0$
to $2\pi$ and the oscillator's position can be changed to decrease
or increase the distance $d$. The solution of the left oscillator
is described in the following way:
\[
\textrm{L}_{pl}^{nl}
\]
where: $nl$ is the number of the attractor (in case of multiple attractors
of isolated oscillator) and $pl$ is the period of given attractor
in respect to the period of excitation (we assume that solutions are
periodic). Similarly, the solution of the right oscillator is given
by:
\[
\textrm{R}_{pr}^{nr}
\]
where: $nr$ is the number of the attractor (in case of multiple attractors
of isolated oscillator), $pr$ is the period of given attractor. To
define the solution of the interacting oscillators system we will
use the following notations:
\[
\textrm{L}_{pl}^{nl}\textrm{R}_{pr-s}^{nr}.
\]
where $s$ is the shift in phase between the systems given by an integer
number when the period of solution is longer than the period of excitation
i.e, $s=1$ for $2\pi$ shift and so on.

The best example to describe the importance of $s$ is a case when
we have two identical systems both with the same period-2 solution
(i.e. their response periods are twice longer than the period of excitation).
In Poincar\'{e} map, for both systems, we observe two dots. Let's assume
that the position of the first oscillator, at the sampling moment
of time, is in one of the dots. Then, the second oscillator can be
either in the same position ($s=0$) or in the second dot when its
phase is shifted by one period of excitation ($s=1$). Number of possible
shifts is equal to the greatest common divisor ($\gcd$) of both systems solutions'
periods. Let us now consider an example where both
oscillators have period-2 and period-5 co-existing solutions.
If the first oscillator is on period-2 solution and the second one
is on period-5 solution only one configuration is possible because
$\gcd\left(2,5\right)=1$, so we have one possible value of $s=0$.
In the other case where both oscillators have period-4 and period-6 co-existing solutions, the $\gcd\left(4,6\right)=2$, hence $s=0$
or $s=1$.

\begin{figure}[H]
\centering{}
\includegraphics{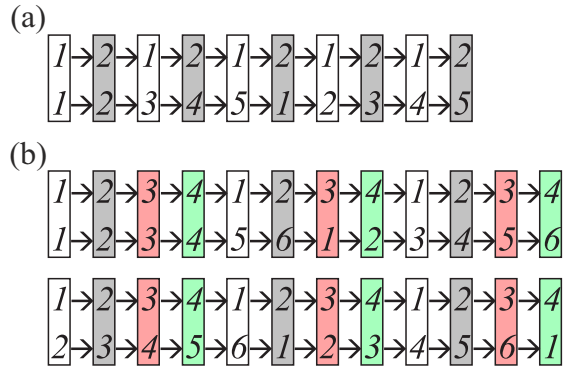} %\includegraphics[bb=0bp 0bp 400bp 123bp,clip]{Number}

\protect\protect\protect\caption{The possible combinations of the system states for (a) period-2 and period-5 solutions and (b) period-4 and period-6.\label{fig:Figure1-2-2}}
\end{figure}

Figure \ref{fig:Figure1-2-2} demonstrates these examples. In Figure
\ref{fig:Figure1-2-2}(a) we show possible configuration for systems
with periods 2 and 5. In this case the period of the whole system is equal
to $10$. The upper row shows the sequence of possible positions of the
system with period 2 (1st or 2nd dot on Poincare map), while the lower
row presents the possible positions of the system with period 5 (1st to
5th dot on Poincare map). It is easy to see that along one period
of the whole system all possible pairs of numbers are created, hence no
more states can occur and any shift makes no difference. In Figure
\ref{fig:Figure1-2-2}(b) we show the second example when two possible
configurations for systems with periods 4 and 6. The first case for $s=0$
(upper) where both system start form dot No. 1 (one in the lower
and the upper row) and the second case for $s=1$ when the first oscillator
starts motion from dot No. 1 and the second system from dot No. 2.
Analysing the graph we see that shifts $s=0$ and $s=1$ create all
possible pairs of dots on Poincare map.

For example, if each oscillator has only one period-2 solution, the
solutions of the interacting system can be described as $\textrm{L}_{2}^{1}\textrm{R}_{2-0}^{1}$
and $\textrm{L}_{2}^{1}\textrm{R}_{2-1}^{1}$. More examples and the
detailed explanations for the possible solutions in the system with
co-existing attractors are given in the following sections.

It should be noted that in our investigations we use the relative
position of two oscillators on their attractors which depend on the
phase shift of excitation $\varphi$. Therefore, in the plots each
presented solution is marked as
\[
\textrm{L}_{pl}^{nl}\textrm{R}_{pr-s}^{nr}(\varphi).
\]

\subsection{Possible non-impacting states of the considered system}

\begin{figure}[H]
\centering{}

\includegraphics{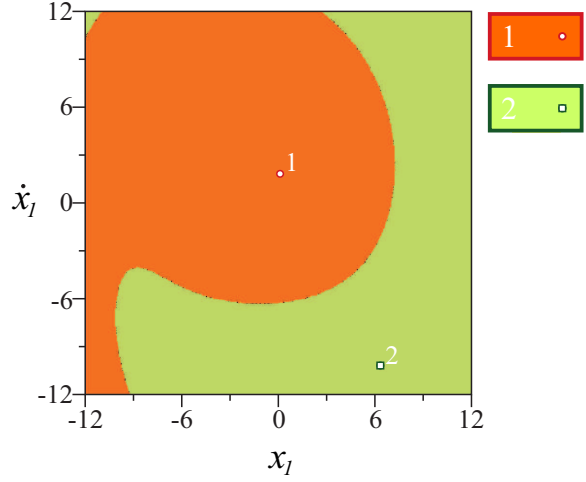}

\protect\protect\protect\caption{Basins of attraction for one Duffing oscillator. ($M=1.0$, $k_{1}=1.0$,
$k_{2}=0.01$, $c=0.05$, $F=1.0$, $\omega=1.3$). \label{fig:Figure1-2}}
\end{figure}

After defining the notations we can continue the analysis of two Duffing
systems introduced at the beginning of this section. For the assumed parameters single Duffing
oscillator has two period-1 solutions
with large and small amplitude, respectively. Their basins of attraction
are shown in Figure \ref{fig:Figure1-2}, where the non-resonant solution with small
amplitude is marked by No. 1 (orange basin) and the resonant solution with
large amplitude by No. 2 (green basin). The basins'
boundary is a smooth curve so it is clear which attractor is reachable
for the given initial condition. All numerical calculations are performed using Runge-Kutta 4th order method with constant time step equal to $T/{1440}$, where $T$ is period of excitation.

\begin{figure}[H]
\centering{}

\includegraphics{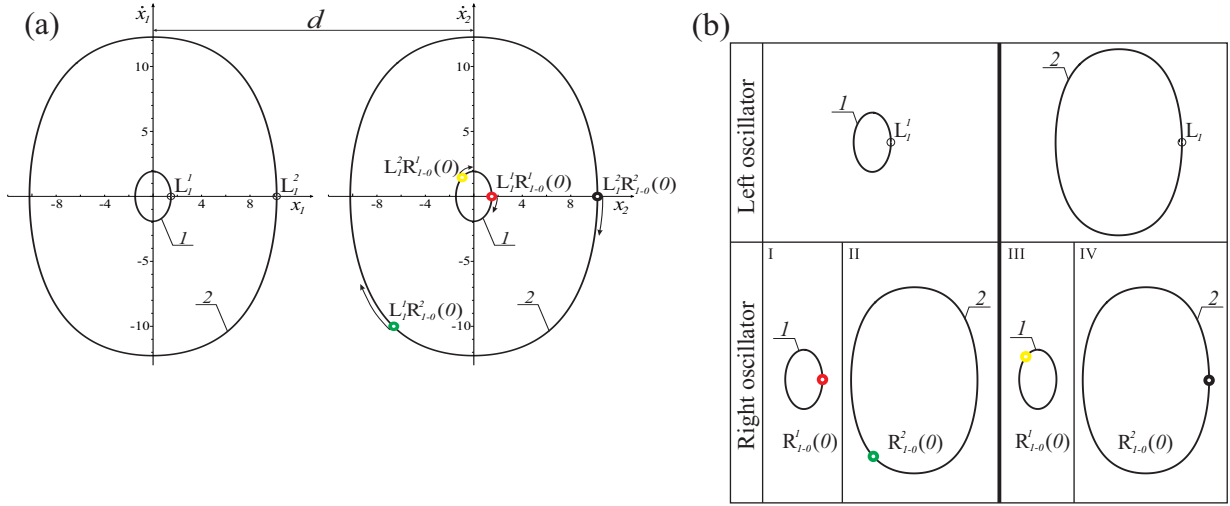}

\protect\protect\protect\caption{Possible attractors of two Duffing oscillators for in-phase forcing
($\varphi=0$) (a) and four possible solutions of the system that
consists of two Duffing oscillators (b). ($M=1.0$, $k_{1}=1.0$,
$k_{2}=0.01$, $c=0.05$, $F=1.0$, $\omega=1.3$). \label{fig:Figure1-2-1}}
\end{figure}

For two identical systems shown in Figure \ref{fig:Figure1}, trajectories
of two coexisting period-1 solutions on the phase plane are presented
in Figure \ref{fig:Figure1-2-1} (a) for the following values of the
system parameters ($M=1.0$, $k_{1}=1.0$, $k_{2}=0.01$, $c=0.05$,
$F=1.0$, $\omega=1.3$). According to the introduced notation, the
solution of the left system can be either the period-1 attractor with
small amplitude (small loop) denoted by $L_{1}^{1}$ or the period-1 attractor with large amplitude (large loop) denoted by $L_{1}^{2}$.
The dots on small and large loops of the left oscillator's trajectories
correspond to the maximum displacements which the oscillator could
reach. When the left system is in the point $L_{1}^{1}$ the solution
of the right system can be either on the small amplitude period-1
attractor represented by the dot on the small loop or on the large
amplitude period-1 attractor (the dot on the large loop) - this corresponds
to solutions $L_{1}^{1}R_{1-0}^{1}(0)$ (red dot) or $L_{1}^{1}R_{1-0}^{2}(0)$
(green dot). When the left system is in the point $L_{1}^{2}$, the
possible locations of the second oscillator are $L_{1}^{2}R_{1-0}^{1}(0)$
(yellow dot) and $L_{1}^{2}R_{1-0}^{2}(0)$ (black dot). Note that
when the systems are on the different trajectories (the left on the
small and the right on the large and vice verse) their positions are
shifted in phase without the phase shift in excitation ($\varphi=0$).
The detailed scheme of solutions is shown in Figure \ref{fig:Figure1-2-1}
(b).

\begin{figure}[H]
\centering{}
\includegraphics{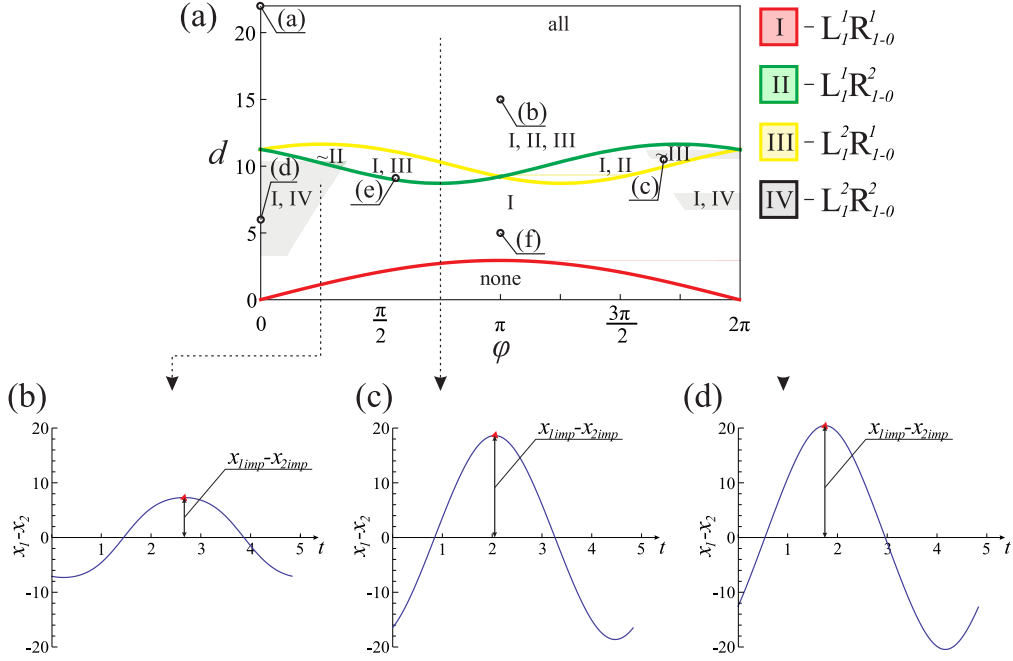} %\includegraphics[bb=0bp 0bp 425bp 280bp,clip]{Duffing_Two_Parameter}

\protect\protect\protect\caption{Possible solutions of the system that consists of two Duffing oscillators
in the $(d,\;\varphi)$ plane (a). Each type of solution is drawn
with line of different color. Subplots (b-d) present trajectory traces
of $L_{1}^{2}R_{1-0}^{2}$ solution in plane ($x_{1}$, $x_{2}$)
for different phase shifts $\varphi$, i.e., (a) $\varphi=\pi/4$;
(b) $\varphi=3\pi/4$; (c) $\varphi=\pi$. Red triangles represent
the collision points. ($M=1.0$, $k_{1}=1.0$, $k_{2}=0.01$, $c=0.05$,
$F=1.0$, $\omega=1.3$). \label{fig:Linear_areas}}
\end{figure}

As we mention before, our controlling parameters are the phase shift
of excitation $\varphi$ and the initial distance between the oscillators.
The current distance between the oscillators can be measured as $x_{1}-x_{2}-d$
and it depends on the relative position of each oscillator along
the trajectory governed by the phase shift $\varphi$. Impact between
the two oscillators occurs when the distance $x_{1}-x_{2}-d$ is zero
or positive. The presence of impacts in the steady state response
introduces new types of solutions in the combined system and depending
on the values of $d$ and $\varphi$ destroys the combinations of
period-1 solutions described earlier. In Figure \ref{fig:Linear_areas}(a)
two parameters plot is presented showing the boundaries for which
given combinations of period-1 solutions exist. Considered system
has four possible combinations of the solutions and each of them is
marked by a different colour. Solution $L_{1}^{1}R_{1-0}^{1}(\varphi)$
disappears for all values of $d$ and $\varphi$ which are located
below the red line on the graph. For $\varphi=0$, the critical distance
$d=d_{crit}$ is equal to zero because the oscillators are moving in phase,
so no collision is possible (they are in contact and $x_{1}=x_{2}$). For $\varphi=\pi$ oscillators move in anti-phase, hence the impact (destruction of given solution) occurs as soon as trajectories start to overlap.

To calculate the boundaries presented in Figure \ref{fig:Linear_areas}(a),
the distance between two oscillators needs to be monitored. In Figures
\ref{fig:Linear_areas}(b-d) the difference between the displacements
of two oscillators $x_{1}-x_{2}$ over one period of steady state
motion for $L_{1}^{2}R_{1-0}^{2}(\varphi)$ solution is shown for
three different cases, i.e, $\varphi=\pi/4$, $\varphi=3\pi/4$, and
$\varphi=\pi$. The red triangle marks indicate the maximum difference
between the displacements during one period of this solution. As the
impact occurs when $x_{1}-x_{2}-d>0$, choosing the critical distances
$d_{crit}$ for given phase shifts $\varphi$ based on these maximum
differences, i.e. (b) $d_{crit}=7.44466$, (c) $d_{crit}=18.66971$
and (d) $d_{crit}=20.40519$, would guarantee that for any $d>d_{crit}$,
there will be no collisions between two oscillators and therefore
this type of solution will exist. For $d<d_{crit}$ the coupled systems
interact and the solution is destroyed. Before oscillators settle down on the attractor we observe the transient dynamics due to non-identical initial conditions. However it does not influence the existence and stability of non-impacting solutions. As can be seen from these graphs, the critical distance $d_{crit}$ is a function of the phase shift $\varphi$ and is valid for given pair of solutions. In other words, given solution disappears when distance between oscillators becomes smaller then $d_{crit}$.

The graph shown in Figure \ref{fig:Linear_areas}(a) gives us a clear
understanding of how the value of the phase shift $\varphi$ influences
the critical distance $d_{crit}$. As one can expect, the solution
that remains stable for the lowest $d$ is the case when two oscillators
are both settled on the small attractors ($L_{1}^{1}R_{1-0}^{1}(\varphi)$).
Such solution exists for all values of $d$ and $\varphi$ above the
red line. Two solutions where one system is on small attractor and
the second one is on large attractor (yellow and green curves) disappear
around $d=10$ and are mirrored in respect to $\varphi=\pi$. The
biggest influence of the shift phase on critical $d_{crit}$ is observed
for solution $L_{1}^{2}R_{1-0}^{2}(\varphi)$ marked by black line. The Roman
numbers printed in each area indicate which types of solutions
are present there. All states are possible above all lines and crossing
each line causes disappearance of one of them, finally below the red
line no non-impacting solution is present. Those lines have been calculated based on trajectories of uncoupled systems and are borders of stability of given solution.

\begin{figure}[p]
\centering{}

\includegraphics{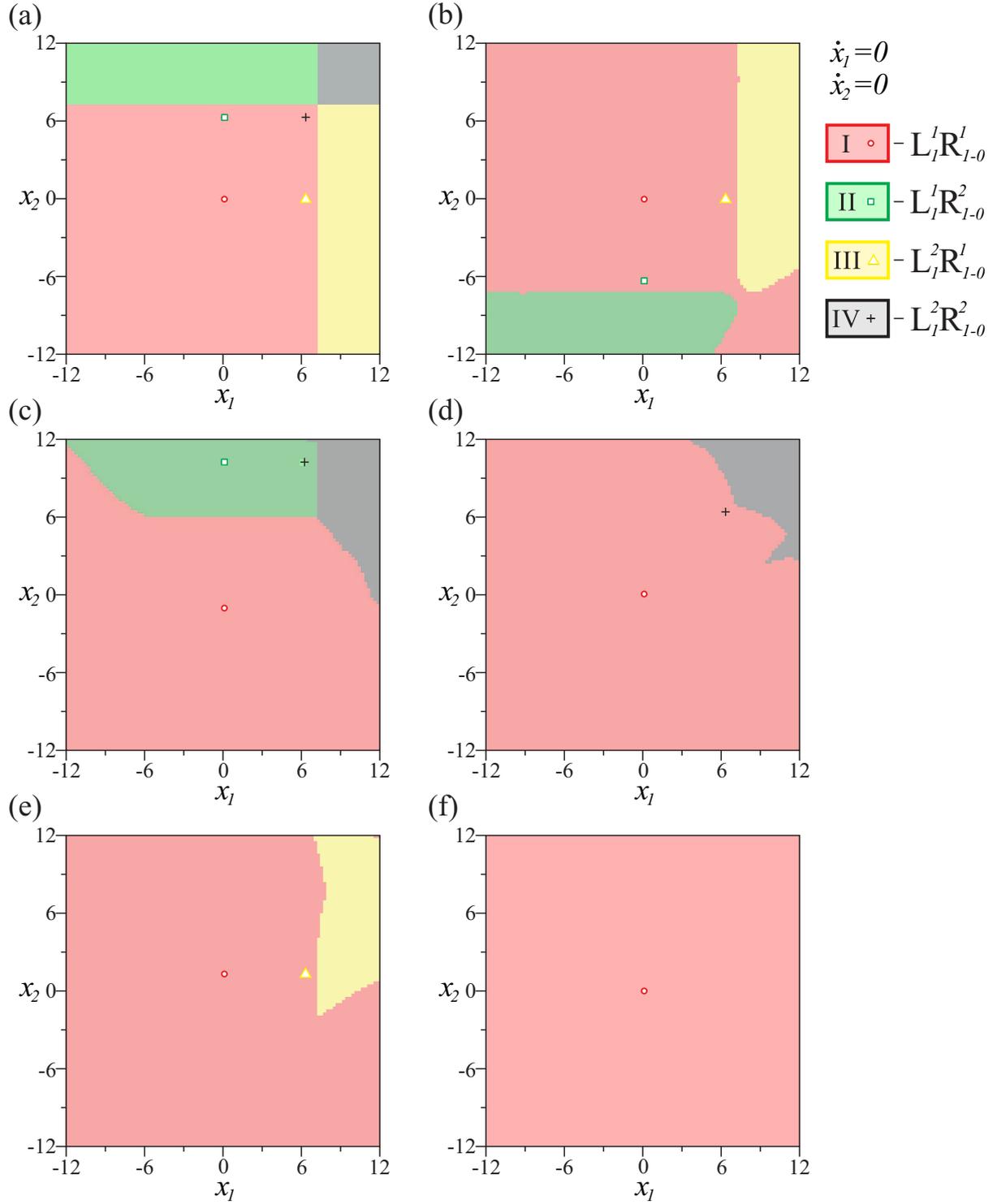}

\protect\protect\protect\caption{Basins of attraction for two Duffing oscillators for the following
values of the system parameters $M=1.0$, $k_{1}=1.0$, $k_{2}=0.01$,
$c=0.05$, $F=1.0$, $\omega=1.3$, $k_{c}=8.0$, $c_{c}=10.0$ for
all subplots except subplot (c) for which $k_{c}=10.0$, $c_{c}=2.0$.
The forcing phase shift and the distance between the oscillators are
the following: (a) $\varphi=0$, $d=22$ (no impacts); (b) $\varphi=\pi$,
$d=15.0$; (c) $\varphi=5.28$, $d=10.5$; (d) $\varphi=0$, $d=6.0$;
(e) $\varphi=0.573\pi$, $d=9$; (f) $\varphi=\pi$, $d=5.0$. Attractors
are marked as dots. Pairs of colours for given attractor and its basin
are shown on right side of the figure. \label{fig:basinLinear}}
\end{figure}

In Figure \ref{fig:basinLinear} we present six basins of attractions
in plane $(x_{1},\: x_{2})$ calculated for the initial
velocities of oscillators fixed to zero, i.e., $\dot{x}_{1}=\dot{x}_{2}=0.0$. As
the full dimension of phase space is four (i.e. the complete set of
initial conditions for the system includes two displacements and two
velocities), it is difficult to visualize the basins of attraction
for all possible initial states. Therefore, two dimensional cross-sections
of the phase space are presented with two most relevant state variables
as parameters. It is clear that in two dimensional projection each
oscillator should be represented by one variable. In mechanical
systems one can easily control and precisely impose the initial displacements,
while such control is impossible or very difficult for the velocities.
Therefore, in the current study the velocities of both systems were
fixed as zero and only displacements are varied.

Each basin is computed for different $\varphi$ and $d$ to show that
varying of the phase and the distance between two oscillators changes
the number of coexisting solutions. Circles in Figure \ref{fig:Linear_areas}(a)
marked by letters (a)-(f) correspond to values of $\varphi$ and $d$
for which the basins are calculated. Figure \ref{fig:basinLinear}(a)
shows basins for two oscillators calculated for in phase forcing ($\varphi=0$)
and $d=22$, the boundaries are straight lines and there is no coupling
between the systems. When we decrease the distance, oscillators start
to interact and one can observe disappearance of one or more solutions.
In Figure \ref{fig:basinLinear}(b) computed for $\varphi=\pi$ and
$d=15.0$, three coexisting solutions are present (Nos. I, II and
III). This corresponds to point (b) in Figure \ref{fig:Linear_areas}(a) (below the black line) in plane
($\varphi$, $d$), hence the
case where two systems are on big loops is destroyed. In Figure \ref{fig:basinLinear}(c)
one can observe three solutions (Nos. I, III and IV) - disappearance
of $L_{1}^{1}R_{1-0}^{2}(5.28)$. Two next plots
correspond to set of parameters for which only two solutions are observed
(see Figure \ref{fig:basinLinear}(d,e)) and the last possibility
with just one present solution is shown in Figure \ref{fig:basinLinear}(f).

It is important to clarify the position of the points marking the
different attractors in the Figure \ref{fig:basinLinear}. As can
be seen from this figure, in some cases the attractors shown do not
belong to their own basin (i.e. yellow, green and black symbols in
Figure \ref{fig:basinLinear}(a)). This can be explained by the fact
that the basins are calculated for the two dimensional projection
$(x_{1},x_{2},0,0)$ and the real Poincare section (i.e. values of
displacements $x_{1}$ and $x_{2}$ at moments $t_{n}=2\pi n/\omega$)
needs to be projected on the chosen plane. The projection of attractor
can be then in any place on the two dimensional cross-section, not
necessarily within its basin of attraction. Nevertheless, we do not
resign form showing attractors because number of dots give information
about the periodicity of the solution.

It is interesting to note that in all examples presented in Figure
\ref{fig:basinLinear}, only non impacting solutions were found and
the number of them were controlled by the phase shift $\varphi$ and the distance between the oscillators $d$. In general, as the distance decreases, the co-existence of the non impacting and impacting solutions becomes likely.
For the small distances (below red line given in Figure \ref{fig:Linear_areas}(a)),
only impacting solution(s) exist.

\subsection{Bifurcation sequence}

From practical point of view it is important to know the bifurcation scenarios of each solution. In this subsection we show how the destruction of non-impacting solutions occur when the distance between the oscillators $d$ decreases for fixed value of $\varphi$ (see Fig. \ref{fig:basinLinear}).  As in the considered system we have four non-impacting solutions we calculate four bifurcation diagrams.  In the  narrow range of $d$ after destruction of non-impacting solutions we observe an existence of impacting solutions with small basins of attractions.

\begin{figure}[H]
\centering{}
\includegraphics{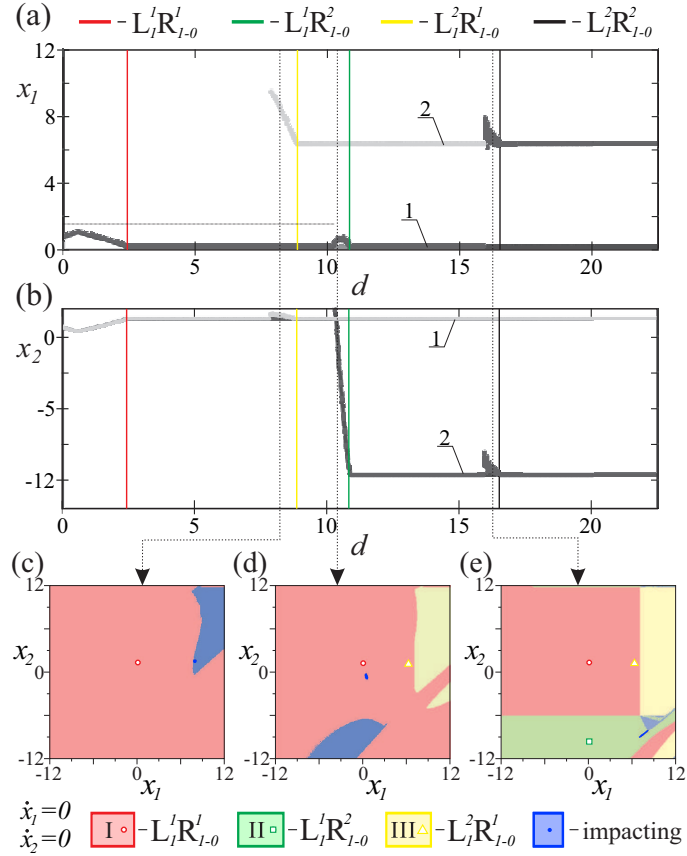} %\includegraphics[bb=0bp 0bp 425bp 280bp,clip]{Duffing_Two_Parameter}

\caption{Bifurcation diagrams of left system $L$ (a) and right system $R$ (b) for four periodic non-impacting solutions versus distance $d$ for $\varphi=5/8\pi$, $c_c=20$ and $k_c=8$. Panles (c-e) show basins of attractions for marked values of distance $d$. i.e, $d=8.2$ (c) $d=10.25$ (d) and $d=16.25$ (e). The blue colour indicates the attractors with impacts.  \label{fig:bif_diag}}
\end{figure}

We fix the phase $\varphi=5/8\pi$, stop parameters $k_c=8.0$ and $c_c=20.0$ and start our computation from $d=22.5$. The results are presented in Figure \ref{fig:bif_diag}(a,b). In Figure \ref{fig:bif_diag}(a) we present the state of the left system ($L$), while in Figure \ref{fig:bif_diag}(b) the state of the right system ($R$). We use vertical lines to mark when the destruction of given solution occurs (colours of lines correspond to the ones used in Figures \ref{fig:Linear_areas} and \ref{fig:basinLinear}). The first solution that is destabilized in grazing bifurcation for $d=16.50$ (black line) is $L_{1}^{2}R_{1-0}^{2}$. When further decreasing $d$ we observe a short transient impacting solution ($d\in [15.96,\:16.5]$) and finally the transition to non-impacting solution $L_{1}^{1}R_{1-0}^{2}$. Hence, then only three stable non-impacting solutions coexists. The second solution $L_{1}^{1}R_{1-0}^{2}$ destabilizes in grazing bifurcation for $d=10.85$ and after transient range it merges with $L_{1}^{1}R_{1-0}^{1}$ for $d=10.2$. Solution $L_{1}^{2}R_{1-0}^{1}$ changes its stability in grazing bifurcation for $d=8.8$ and merges with $L_{1}^{1}R_{1-0}^{1}$ for $d=7.85$.
The last non-impacting solution ($L_{1}^{1}R_{1-0}^{1}$) disappears, similarly to the mentioned before, in grazing bifurcation for $d=2.45$ and we can observe only attractor with impacts. For all non-impacting solutions the destabilization is followed by the appearance of impacting solution which disappears when further decreasing the distance $d$.

In Figure \ref{fig:bif_diag}(c-e) we show three basins of attractions calculated after the destabilization of non-impacting solutions for $d=8.2$ (c) $d=10.25$ (d) and $d=16.25$ (e). We hold the same colours of basins of attraction as in Figure \ref{fig:basinLinear}; additionally the blue colour indicates the basins of attraction of impacting solutions. In all three panels area of the blue basin is small comparing to the total area of the plot, i.e., panel (c) - $6.3 \%$, panel (d) - $5.8 \%$ and panel (e) - $0.8 \%$.

Although we are not able to completely avoid impacting solutions, they are only present for some small ranges of $d$ close to destabilization of non-impacting solution and these range can be always decreased with proper tuning of stop parameters $k_c$ and $c_c$.
Moreover, even if the impacting solutions coexists their basins of attraction are not dominant.

\section{Bi-linear oscillator}

\begin{figure}[H]
\centering{}
\includegraphics{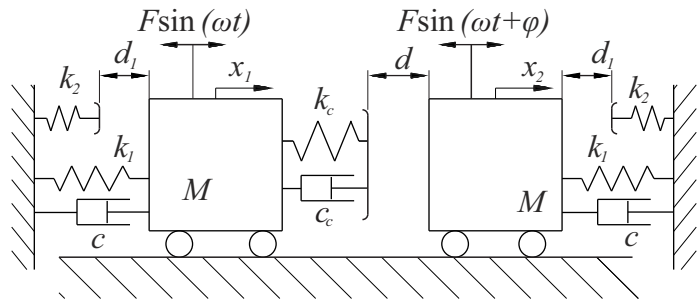} %\includegraphics[bb=0bp 0bp 370bp 90bp,clip]{Katya_model}

\protect\protect\protect\caption{Model of two discontinuously coupled bi-linear oscillators. \label{fig:bilinear_model} }
\end{figure}

In this section we show that the behaviour presented for the system
of two colliding Duffing oscillators is common for systems with impacts.
Specifically, here it is demonstrated that with the properly
chosen distance between the oscillators and difference in phase of
the harmonic excitation, we can reduce the number of possible solutions.
The same phenomenon was also observed in the coupled bi-linear systems
shown in Figure \ref{fig:bilinear_model}. As in the previous case, when two oscillators remain at rest,
they are separated by distance $d$, and the impacts which
occur between them are of the soft type because of the presence of
the spring with stiffness $k_{c}$ and the viscous damper with damping
coefficient $c_{c}$. Both oscillators have masses $M$ and they are
supported by the viscous dampers with damping coefficient $c$ and
by two linear springs of stiffnesses $k_{1}$ and $k_{2}$ which provide
piecewise linear elastic resistance force. Springs with stiffness $k_{2}$ are separated from masses $M$ by distance $d_1$.  Similarly to the Duffing
systems, both oscillators are excited by harmonic force with amplitude
$F$ and frequency $\omega$. Forcing of the first oscillator has
fixed phase shift (equal to zero) while for the second one the phase
shift is $\varphi\in\left\langle 0,\:2\pi\right)$ which is the control
parameter of the coupled system.

The equations of motion for this coupled system are:

\begin{equation}
M\ddot{x}_{1}+k_{1}x_{1}+k_{2}\left(x_{1}+d_{1}\right)H\left(-x_{1}-d_{1}\right)+c\dot{x}_{1}+F_{C}=F\sin\left(\omega t\right)
\label{eq:Bilin_1-2}
\end{equation}

\begin{equation}
M\ddot{x}_{2}+k_{1}x_{2}+k_{2}\left(x_{2}-d_{1}\right)H\left(x_{2}-d_{1}\right)+c\dot{x}_{2}-F_{C}=F\sin\left(\omega t+\varphi\right)
\label{eq:Bilin_1-1-2}
\end{equation}
where$H(\cdot)$ is a Heaviside step function. Force generated by
the discontinuous coupling ($F_{C}$) is given by the formula:

\begin{equation}
F_{C}=\begin{cases}
\begin{array}{cc}
0\\
\\
k_{c}\left(\left(x_{1}-x_{2}\right)-d\right)+c_{c}\left(\dot{x}_{1}-\dot{x}_{2}\right)
\end{array} & \begin{array}{cc}
for & x_{1}-x_{2}<d\\
\\
for & x_{1}-x_{2}\geqslant d
\end{array}\end{cases}
\label{eq:bilin_1-1-1-1}
\end{equation}
The parameters have the following values: $M=1.0\,[kg]$, $k_{1}=1.0\,[\frac{N}{m}]$,
$k_{2}=29.0\,[\frac{N}{m}]$, $c=0.02\,[\frac{Ns}{m}]$, $F=0.518\,[N]$,
$\omega=0.86\,[\frac{1}{s}]$, $d_{1}=1.26\,[m]$, $k_{c}=8.0\,[\frac{N}{m}]$,
$c_{c}=10.0\,[\frac{N}{m}]$, $d=1.0\,[m]$.

Introducing non-dimensional time $\tau=t\omega_{1}$, where $\omega_{1}=1\,[\frac{1}{s}]$,
reference length $l_{r}=1.0\,[m]$ and mass $m_{r}=1\,[kg]$ we transform
the equations (\ref{eq:Bilin_1-2})  -- (\ref{eq:bilin_1-1-1-1})
into non-dimensional form in which dimensional parameters
are replaced by the following non-dimensional parameters:

\begin{center}
$M^{\prime}=\frac{M}{m_{r}}$,
$k_{1}^{\prime}=\frac{k_{1}l_{r}}{m_{r}\omega_{1}^{2}}$, $k_{2}^{\prime}=\frac{k_{2}l_{r}}{m_{r}\omega_{1}^{2}}$,
$c^{\prime}=\frac{c}{m_{r}\omega_{1}}$, $F^{\prime}=\frac{F}{m_{r}l_{r}\omega_{1}^{2}}$,
$\omega^{\prime}=\frac{\omega}{\omega_{1}}$, $d_{1}^{\prime}=\frac{d_{1}}{l_{r}}$,
$k_{c}^{\prime}=\frac{k_{c}l_{r}}{m_{r}\omega_{1}^{2}}$, $c_{c}^{\prime}=\frac{c_{c}}{m_{r}\omega_{1}}$,
$d^{\prime}=\frac{d}{l_{r}}$.
\end{center}
Transformation to the dimensionless
form was performed in the way that enables to hold the parameters'
values, hence: $M^{\prime}=1.0$, $k_{1}^{\prime}=1.0$, $k_{2}^{\prime}=29.0$,
$c^{\prime}=0.02$, $F^{\prime}=0.518$, $\omega^{\prime}=0.86$,
$d_{1}^{\prime}=1.26$, $k_{c}^{\prime}=8.0$, $c_{c}^{\prime}=10.0$. Distance between system $d$ and phase shift in excitation $\varphi$ are controlling parameters. For simplicity all of the primes used in definitions
of dimensionless parameters will be omitted hereafter in the analysis.

\begin{figure}[H]
\centering{}
\includegraphics{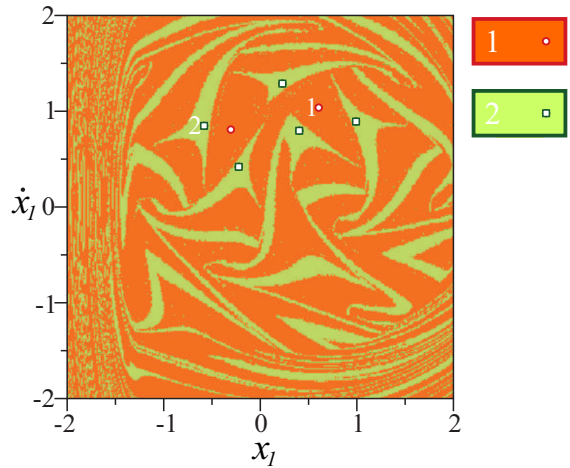} %\includegraphics[bb=0bp 0bp 225bp 175bp,clip]{Katya_s_Basins_one}

\protect\protect\protect\caption{Basins of attraction for one bi-linear oscillator for $\omega=0.86$
and $F=0.518$. Two solutions are observed: with period 2 and period 5.
($M=1.0$, $k_{1}=1.0$, $k_{2}=29.0$, $c=0.02,$ $d_{1}=1.26$)
\label{fig:Basin_single_bilinear}}
\end{figure}

It is well known that a single bi-linear oscillator exhibits a complex
non-linear behavior \cite{pavlovskaia2010complex} and the co-existing
attractors are widespread phenomena in this system. A typical example
of the basins of attractions for two co-existing solutions of a single
bi-linear oscillator is presented in Figure \ref{fig:Basin_single_bilinear}
for $\omega=0.86$ and $F=0.518$. Here we observe two coexisting
attractors; the first one is a period-2 response which period is twice
longer than the period of the excitation. It is marked by red dots
and its basin is given in orange colour. The second one is a period-5 response which is denoted by dark green squares and its basin is in
green colour. Figure \ref{fig:Basin_single_bilinear} displays that the basin of the period-2 solution
is dominant.

\begin{figure}[h]
\centering{}
\includegraphics{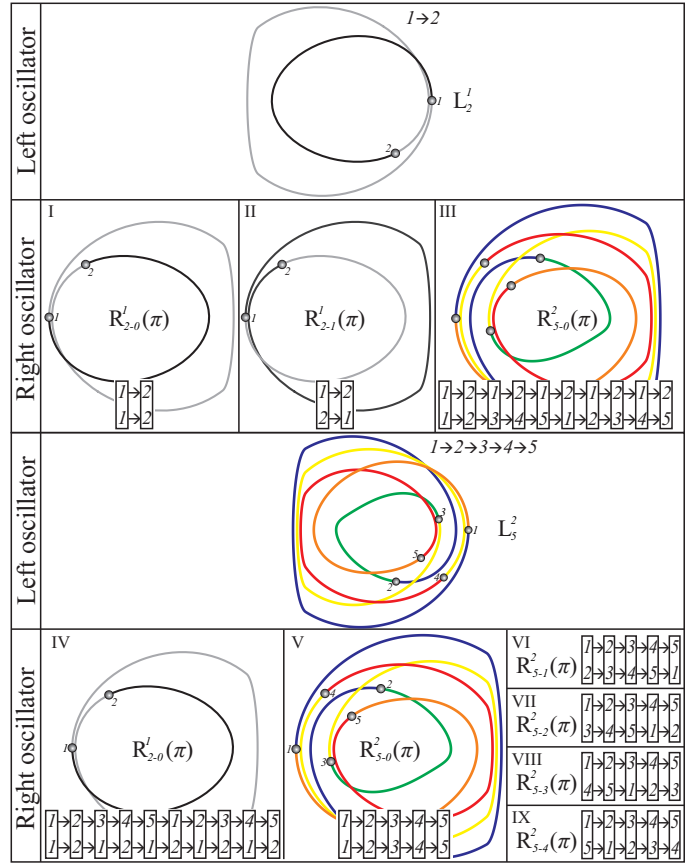} %\includegraphics[bb=0bp 0bp 495bp 165bp,clip]{Katya_s_Possibilities}

\protect\protect\protect\caption{Nine possible solutions of the system that consists of two bi-linear
oscillators. ($M=1.0$, $k_{1}=1.0$, $k_{2}=29.0$, $c=0.02,$ $d_{1}=1.26$,
$\omega=0.86$, $F=0.518$) \label{fig:BiLin_fir2}}
\end{figure}

Similarly as in the system of coupled Duffing oscillators, when we couple two bi-linear oscillators which have two co-existing solutions each, we observe many coexisting periodic states of the coupled system for large distant $d$.
Let's fix the first system on
the period-2 attractor. In such a case the second system has three
possible states shown in Figure \ref{fig:BiLin_fir2}. The gray dots
correspond to position of system at starting point and after the full
period of excitation; the upper window and windows I and II
the first part of the trajectory is marked by black line and the second
part is by grey line. In the first and the second solution ($L_{2}^{1}R_{2-0}^{1}$
and $L_{2}^{1}R_{2-1}^{1}$) both systems have period-2 responses. Hence,
we can observe zero ($s=0$) or one period of excitation shift ($s=1$).
In the third solution $L_{2}^{1}R_{5-0}^{2}$ the second oscillator
is on the period-5 attractor, hence the whole system has period 10
(smallest common denominator). As it easy to see, the following
sequence of numbers in panel III in Figure \ref{fig:BiLin_fir2} is
the only possible solution because all possible pairs of numbers are
present in the sequence (for details see Figure \ref{fig:Figure1-2-2}(a)).

When the first system is on the period-5 attractor, we observe six
possible solutions. The first one is a mirror solution of the one
shown in the panel III in Figure \ref{fig:BiLin_fir2}, hence the second
system exhibits the period-2 response. Next five solutions correspond
to the case when both systems are on period-5 attractors, and as can be
seen in panels V-IX in Figure \ref{fig:BiLin_fir2}, the phase of
the second system solution can be either identical to the first one
($s=0$, panel V), or shifted by $2\pi$ ($s=1$, panel VI), $4\pi$
and so on up to $s=4$ (panel IX). Therefore, for the considered system
of two coupled bi-linear oscillators nine different non-impacting states are obtained.

\begin{figure}[h]
\centering{}
\includegraphics{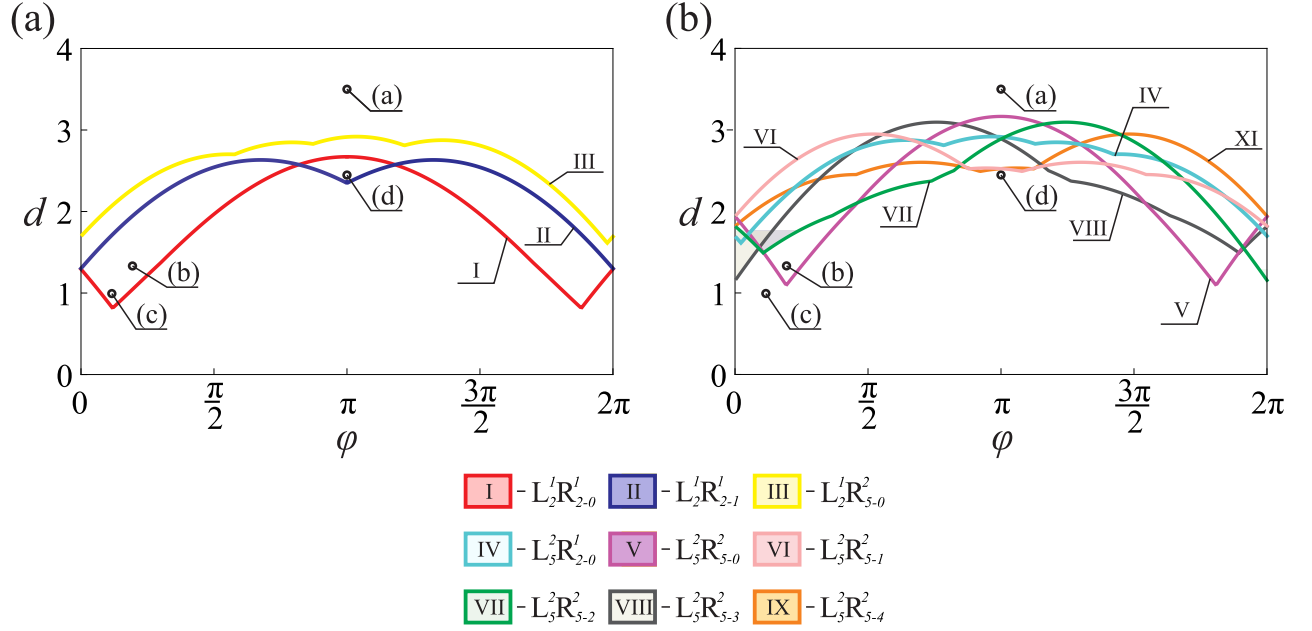} %\includegraphics[bb=0bp 0bp 500bp 145bp,clip]{Katya_s_result}

\protect\protect\protect\caption{Possible solutions of the system that consists of two bi-linear oscillators
in the $(d,\;\varphi)$ plane. Each type of solution is marked and drawn with
line of different color. ($M=1.0$, $k_{1}=1.0$, $k_{2}=29.0$, $c=0.02,$
$d_{1}=1.26$, $\omega=0.86$, $F=0.518$) \label{fig:BiLin_lines}}
\end{figure}

Each of the above mentioned periodic states can be eliminated by the
proper choice of the distance $d$ and the phase shift of the excitation
$\varphi$. The analysis shown in Figure \ref{fig:BiLin_lines} is
presented in two parts for the clarity. In Figure \ref{fig:BiLin_lines}(a)
the red, blue and yellow lines correspond to following solutions:
$L_{2}^{1}R_{2-0}^{1}$, $L_{2}^{1}R_{2-1}^{1}$ and $L_{2}^{1}R_{5-1}^{2}$
and in Figure \ref{fig:BiLin_lines}(b) the light blue, purple, pink,
green, grey and orange lines to $L_{5}^{2}R_{2-0}^{1}$, $L_{5}^{2}R_{5-0}^{2}$,
$L_{5}^{2}R_{5-1}^{2}$, $L_{5}^{2}R_{5-2}^{2}$, $L_{5}^{2}R_{5-3}^{2}$
and $L_{5}^{2}R_{5-4}^{2}$ respectively. Below all those lines we
do not observe a given periodic non-impacting solution.

\begin{figure}[H]
\centering{}
\includegraphics{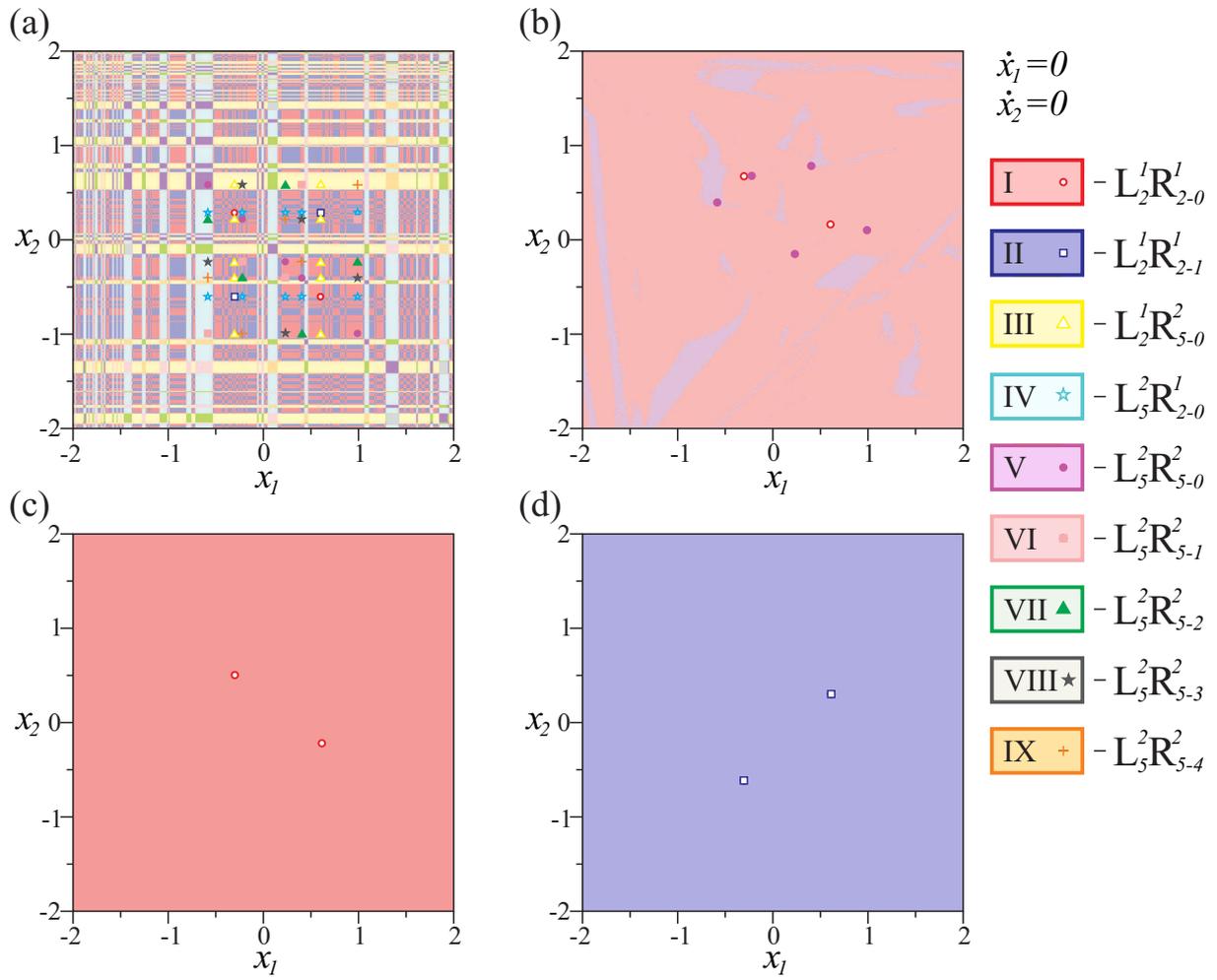} %\includegraphics[bb=0bp 0bp 455bp 370bp,clip]{Katya_s_Basins}

\protect\protect\protect\caption{Basins of attraction for two bi-linear oscillators for $\omega=0.86$,
$F=0.518$ and the following values of the system parameters $M=1.0$,
$k_{1}=1.0$, $k_{2}=29.0$, $c=0.02$, $d_{1}=1.26$, $k_{c}=8.0$,
$c_{c}=10.0$ for all subplots except subplot (d) for which $c_{c}=2.0$.
The forcing phase shift and the distance between the oscillators are
the following: (a) $d=10$ and $\varphi=\pi$ (no impacts); (b) $d=1.332$
and $\varphi=0.60916$; (c) $d=1$ and $\varphi=0.36424$; (d) $d=2.448$
and $\varphi=\pi$. Attractors are marked as dots. Pairs of colours
for given attractor and its basin are shown on right side of the figure.
\label{fig:BiLin_basin_two}}
\end{figure}

In Figure \ref{fig:BiLin_basin_two} four basins of attractions are
presented for different values of $d$ and $\varphi$. Basins in panel
(a) correspond to point (a) shown in Figure \ref{fig:BiLin_lines}(a,b)
with $d=3.5$ and $\varphi=\pi$. In that case all possible non-impacting solutions
are present in the system. In panel (b) (see (b) letter in Figure
\ref{fig:BiLin_lines}(a,b)) we have only two solutions $L_{5}^{2}R_{2-0}^{1}$
and $L_{5}^{2}R_{5-0}^{2}$. Two last panels are calculated for the
pairs of $d$ and $\varphi$ for which only one state is present,
i.e, (c) - $L_{5}^{2}R_{2-0}^{1}$ and (d) $L_{5}^{2}R_{5-0}^{2}$.

\begin{figure}[H]
\centering
\includegraphics{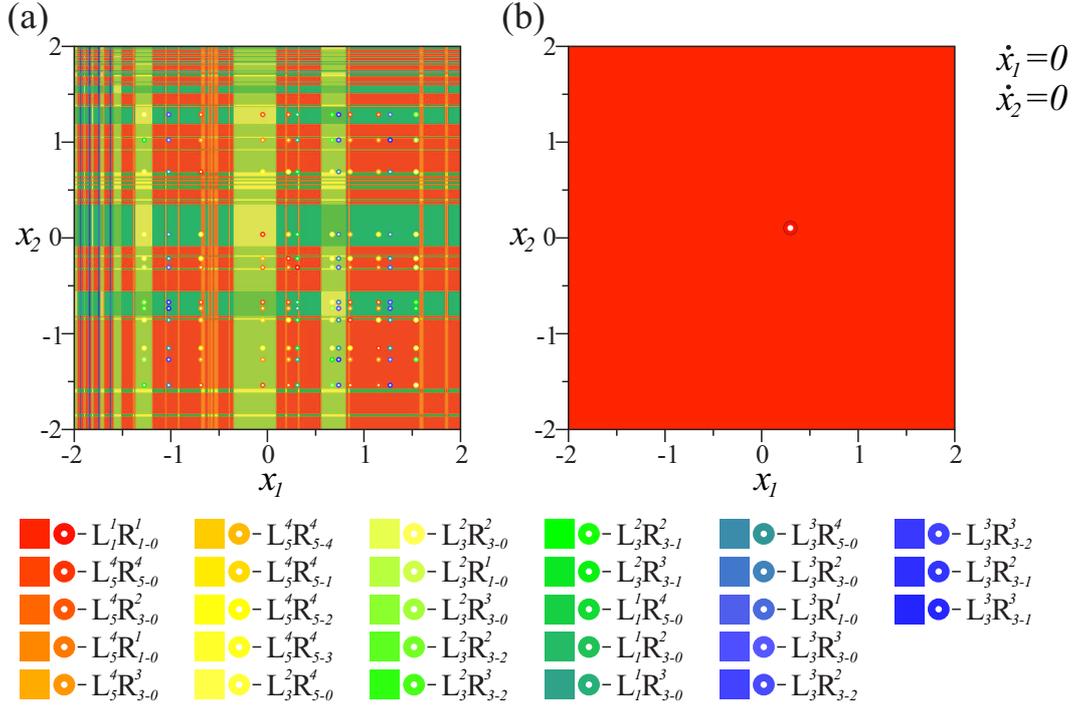} %\includegraphics[bb=0bp 0bp 400bp 265bp,clip]{Katya_s_Basins_2}

\protect\protect\protect\caption{Basins of attraction for two bi-linear oscillators for for $\omega=0.928$
and $F=0.602$ and the following values of the system parameters $M=1.0$,
$k_{1}=1.0$, $k_{2}=29.0$, $c=0.02$, $d_{1}=1.26$, $k_{c}=6.0$,
$c_{c}=5.0$. For given values of parameters single bi-linear oscillator
has four coexisting attractors i.e., one period-1, two period-3
and one period-5. The forcing phase shift and the distance between
the oscillators are the following: (a) $d=10$ and $\varphi=\pi$
(no impacts) and (b) $d=2$ and $\varphi=0.4$. Attractors are marked
as dots. Pairs of colours for given attractor and its basin are shown
bellow the plots. \label{fig:BiLin_basin_four}}
\end{figure}

The last example is shown in Figure \ref{fig:BiLin_basin_four}. We change
the frequency of excitation of bi-linear system to $\omega=0.928$
and $F=0.603$, which in the case of a single bi-linear oscillator
gives us the coexistence of four different attractors, i.e., one with period
one, two with period three and one with period five. For two discontinuously
coupled oscillators at this frequency we observe $28$ possible no-impacting
solutions (see Figure \ref{fig:BiLin_basin_four}(a)). As in the previous
example with decreasing distance $d$ and changing phase $\varphi$
we can reduce the number of solutions. We do not show the whole analysis
because it is similar to the one conducted in the previous cases. We focus on showing
that by choosing two controlling parameters properly the number of
solutions can be reduced to one as shown in Figure \ref{fig:BiLin_basin_four}(b).

\section{Coexistence of impacting and non-impacting solutions}

\begin{figure}[H]
\centering
\includegraphics{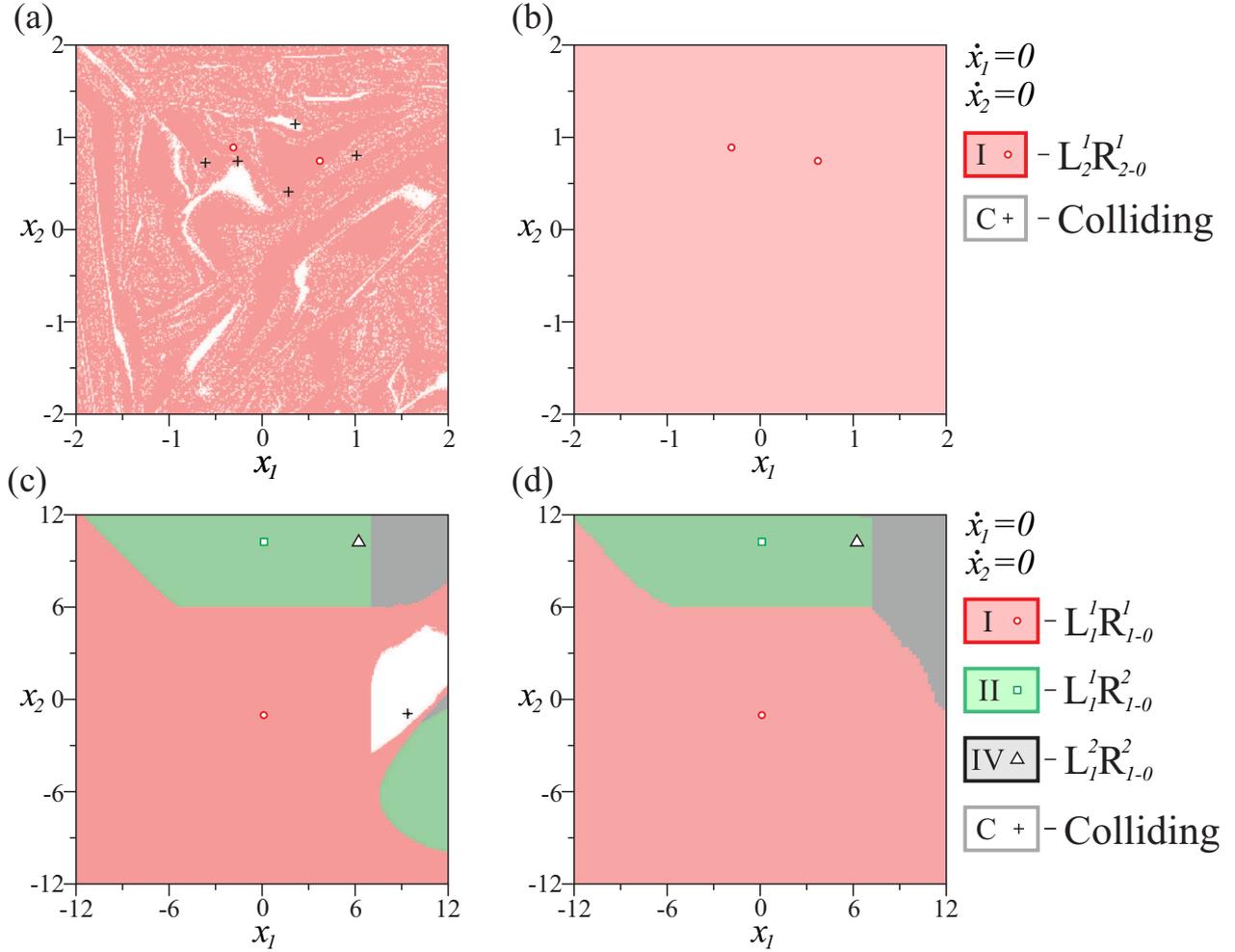} %\includegraphics[bb=0bp 0bp 480bp 370bp,clip]{Colliding_basins}

\protect\protect\protect\caption{Basins of attraction for two bi-linear oscillators (a,b) and two Duffing
oscillators (c,d). Subplots (a,b) were both calculated for the same
system's parameters ($\omega=0.86$, $F=0.518$, $M=1.0$, $k_{1}=1.0$,
$k_{2}=29.0$, $c=0.02$, $d_{1}=1.26$, $\varphi=1.0$, $d=1.5$)
but different coupling parameters $k_{c}=8.0$, $c_{c}=10.0$ for
subplot (a) and $k_{c}=20.0$, $c_{c}=10.0$ for subplot (b). Similarly
subplots (c,d) were calculated for identical system's parameters ($\omega=1.3$,
$F=1.0$, $M=1.0$, $k_{1}=1.0$, $k_{2}=0.01$, $c=0.05$, $\varphi=5.28$,
$d=10.5$) and different coupling parameters: $k_{c}=8.0$, $c_{c}=10.0$
for subplot (c) and $k_{c}=8.0$, $c_{c}=2.0$ for subplot (d). Attractors
are marked as dots. Pairs of colours for given attractor and its basin
are shown on the left side of the figure. \label{fig:Impacting_basins}}
\end{figure}

In this section we show the coexistence of impacting and non-impacting
solutions. The appearance of impacting solutions is probable close
to the border where non-impacting solutions disappear. Such scenario
occurs both for coupled bi-linear and Duffing systems. Near the threshold,
for given $d$ and $\varphi$ one can observe the coexistence of impacting
and non-impacting solutions but such case can be avoided by changing
the parameters of the coupling which is shown in Fig. \ref{fig:Impacting_basins}.
In subplots (a) and (c) one can observe the coexistence of impacting
and non-impacting attractors. For subplots (b,d) we changed coupling
parameters so that only non-impacting solutions exist in the system.
In part (b) we increase $k_c$ from 8.0 to 20 and in part (d) we decrease $c_c$ from 10 to 2.

From the practical point of view the crucial information is the knowledge
how common is the coexistence of impacting and non-impacting solutions.
The analysed systems are complex, hence we can use only numerical
tools to predict how large is the basin of attraction of non-impacting
solutions. To calculate this we use the method proposed by Menck et
al. \cite{menck2013basin}. The idea behind it is simple, however
it is a powerful tool to estimate the size of complex basins of attraction
in multidimensional systems. Let us assume that we want to estimate the volume of basin of attraction {$\mathcal{B}$} of given attractor. To do it we measure the basin stability as $\mathcal{S}_{\mathcal{B}\cap\mathcal{Q}} = Vol{\mathcal{B}\cap\mathcal{Q}}/Vol{\mathcal{Q}}=[0,\:1]$, where $\mathcal{Q}$ is the subset of the state space that has finite volume. Thus, the system is integrated for $N$ initial conditions drawn uniformly at random from $\mathcal{Q}$. Then, the number $M$ of initial condition leading to the expected attractor is summarised and $\mathcal{S}_{\mathcal{B}\cap\mathcal{Q}} = M/N$. Number $N$ in  \cite{menck2013basin} is $500$, in our case due to low dimensionality
of phase space and discontinuity in system's equations we increase
the number of trails to $1000$. For each set of initial values we
check the type of final attractor. Based on this the percentage distribution
of solutions is determined. We want to estimate the participation
of impacting solution, hence we do not distinguish their types but
classify them all as the impacting ones. However, we calculate
the percentage of each non-impacting solution. The ranges of initial
conditions' values are taken in the way to ensure that all considered
attractors are within this multidimensional space, hence none of the attractors
is predominant \cite{chudzik2011multistability,brzeski2016}. Hence they are drawn from wide ranges and none of them have fixed value.  In our calculations
we want to show that the proper choice of coupling stiffness $k_{c}$
or damping $c_{c}$ results in elimination of impacting solutions
(for the other parameters of the systems see captions).

\begin{figure}[H]
\centering
\includegraphics{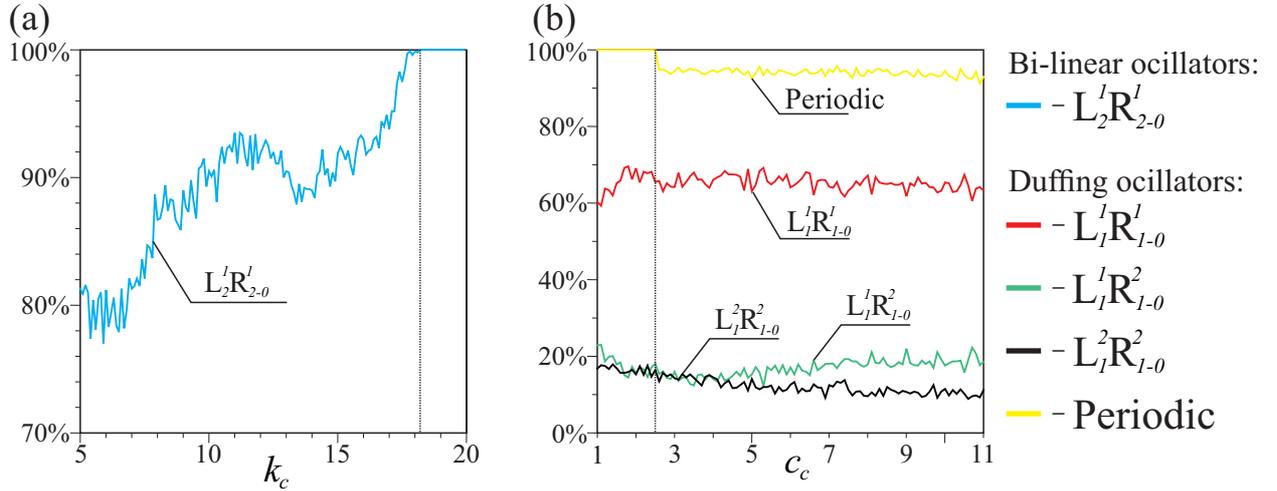} %\includegraphics[bb=0bp 0bp 480bp 370bp,clip]{Colliding_basins}

\protect\protect\protect\caption{Changes of probability of reaching given attractor with the change
of $k_{c}$ parameter for bi-linear oscillators (a), and parameter
$c_{c}$ for Duffing oscillators (b). Colour of lines in subplots (a,b)
correspond to the colours of the attractors presented in Fig. \ref{fig:Impacting_basins}. \label{fig:prob}}
\end{figure}

In Fig. \ref{fig:prob}(a) we show the plot for coupled
bi-linear systems. For bi-linear systems the initial conditions ($x_{1,2}$,
$\dot{x}_{1,2}$) are drawn from the range: $\left[-2,\,2\right]$ and $k_{c}\in\left[5,\,20\right]$.
Values of $d$, $\varphi$ and $c_{c}$ are the same as in Fig. \ref{fig:Impacting_basins}(a,b)
and for this set of parameters only one non-impacting attractor exists. In the
whole considered range of $k_{c}$ the probability of occurrence of
non-impacting solution is over $75\%$ and with increasing of coupling
stiffness it grows up to $100\%$ for $k_{c}>18.2$. The fluctuations
along the probability curve are small (around $2\%$) and typical
for complex, discontinuous systems with high sensitivity on initial
conditions.

The same analysis is performed for coupled Duffing systems (see Fig.
\ref{fig:prob}(b)). Here, all the initial conditions
($x_{1,2}$, $\dot{x}_{1,2}$) are drawn from range: $\left[-12,\,12\right]$
and $c_{c}\in\left[1,\,11\right]$. Values of $d$, $\varphi$ and
$k_{c}$ are the same as in Fig. \ref{fig:Impacting_basins}(c,d).
For this set of parameters three non-impacting solutions exist, hence
we show probability of each of them (colour of the line corresponds
to the colour of the attractor on the basins plot). Moreover we calculate their sum to
show the overall probability of reaching non-impacting solutions (yellow
line). In case of Duffing systems the probability of reaching impacting solutions
is low and does not exceed $9\%$. For low values of $c_{c}<2.5$
the impacting solutions do not exist. The chance of achieving one
of three non-impacting solutions stays nearly constant in the whole
range of $c_{c}$.

The method presented above let us confirm the usefulness of the basins
of attraction calculated for two-dimensional cross-sections of the
four-dimensional phase space (presented in Figs. \ref{fig:basinLinear},
\ref{fig:BiLin_basin_two}, \ref{fig:BiLin_basin_four}, \ref{fig:Impacting_basins}).
To authenticate that, we compare the probability of reaching given attractor obtained using both methods.
For bi-linear oscillators we observe the full agreement. In Fig. \ref{fig:Impacting_basins}(a),
calculated using Dynamics, the chance to reach periodic attractor
is $86\%$, while using Menck et al. method we obtain $85\%$ chance.
For parameters values used in Fig. \ref{fig:Impacting_basins}(b)
both methods give $100\%$ chance of reaching periodic non-impacting
solution (the only stable solutions in the whole phase space). For
Duffing oscillators the difference in the results obtained
using both methods is noticeable. For Fig \ref{fig:Impacting_basins}(c) we have
$69.5\%,$ $21\%$, $5\%$ and $4.5\%$ chance for reaching $L_{1}^{1}R_{1-0}^{1}$,
$L_{1}^{1}R_{1-0}^{2}$, $L_{1}^{2}R_{1-0}^{2}$ and colliding solution
respectively. Using Menck et. al. method we get $65.4\%,$ $17.6\%$,
$11.5\%$ and $5.5\%$. Similar difference between both methods can
be observed for Fig. \ref{fig:Impacting_basins}(d): for solutions
$L_{1}^{1}R_{1-0}^{1}$, $L_{1}^{1}R_{1-0}^{2}$, $L_{1}^{2}R_{1-0}^{2}$
we get $75\%,$ $17.5\%$, $7.5\%$ chances using Dynamics, and $69.1\%,$
$15.6\%$, $15.3\%$ using Menck et. al. method. Hence, for Duffing
oscillators we have up to $8\%$ difference in the results obtained
using both methods.

The good agreement of both methods for bi-linear systems is the result of the
properties of basins of attraction of the single system. As can one see
in Fig. \ref{fig:Basin_single_bilinear} basin of attraction has a
fractal structure, so with changing velocity there is no qualitative
change of probability that we reach given attractor. For Duffing systems
the basins of attraction for single oscillator has regular structure
with sharp borders (see Fig. \ref{fig:Figure1-2}), hence with changing
the velocity the chance of approaching given solution varies significantly
(from $0\%$ to $100\%$). This shows the main advantage of Menck et.
al. method, where the initial conditions are generated from the whole
accessible phase space.

\section{Conclusions }

In this paper we study the influence of discontinuous coupling on the dynamics of multistable systems. It is shown that with precise selection of the systems' parameters
(the gap between the systems or/and phase shift of external excitation)
one can easily decrease the number of coexisting solutions via discontinuous
coupling. This idea is verified by performing calculation for two types
of oscillators. The first system is composed of two Duffing oscillators,
while the second one consists of two bi-linear oscillators. When systems
are uncoupled ($\forall t\;\left(x_{1}-x_{2}\right)<d$) there are
numerous coexisting non-impacting solutions. When oscillators start
to interact we observe impacts. Due to the damper in the stop the
energy is dissipated and oscillators settle down on one of the non-impacting
solutions and perform the synchronous motion (phase
synchronization). The exchange of the energy (impacts) is a transient
state lasting up to some minimal distance $d$ for which only impacting solutions
are presented. Near the threshold where non-impacting
solutions destabilize, one can observe the coexistence of impacting
and non-impacting solutions. However, by the exact selection of stop's
parameters we are able to eliminate the impacting solutions. The proposed method
is robust and the idea is likely to be applied in gear transmission systems, energy harvesters and in control and danger elimination of impacts between densely located oscillators or structures.

\section*{Acknowledgment}

This work has been supported by Lodz University of Technology own
Scholarship Fund (PB) and by Stipend for Young Outstanding Scientists
from Ministry of Science and Higher Education of Poland (PP).

PB is supported by the Foundation for Polish Science (FNP).

%\section*{References}


\begin{thebibliography}{30}

\bibitem{moreau1988nonsmooth}
J.~J.~Moreau and P.~D.~Panagiotopoulos
\newblock \emph{Nonsmooth Mechanics and Its Applications},
\newblock Springer Science \& Business Media, Udine, 1988.


\bibitem{gilardi2002literature}
G.~Gilardi and I.~Sharf.
\newblock Literature survey of contact dynamics modelling
\newblock \emph{Mechanism and Machine Theory} \textbf{37}, 1213 (2002).

\bibitem{brogliato1999nonsmooth}
B.~Brogliato.
\newblock \emph{Nonsmooth mechanics: Models, dynamics and control},
\newblock Springer Science \& Business Media, London, 1999


\bibitem{Hutzler2004}
S.~Hutzler, G.~Delaney, D.~Weaire and F.~MacLeod.
\newblock Rocking Newton's cradle.
\newblock \emph{American Journal of Physics} \textbf{72}, 1508 (2004).

\bibitem{blazejczyk2010hard}
B.~Blazejczyk-Okolewska, K.~Czolczynski, and T.~Kapitaniak. 
\newblock Hard versus soft impacts in oscillatory systems modeling. 
\newblock \emph{Communications in Nonlinear Science and Numerical Simulation},
  \textbf{15}, 1358 (2010).

\bibitem{kundu2011vanishing}
S.~Kundu, S.~Banerjee, and D.~Giaouris. 
\newblock Vanishing singularity in hard impacting systems. 
\newblock \emph{Discrete and Continuous Dynamical Systems - Series B} \textbf{16}  319 (2011).

\bibitem{witelski2014driven}
T.~Witelski, L.~Virgin, and C.~George.
\newblock A driven system of impacting pendulums: Experiments and simulations
\newblock \emph{Journal of Sound and Vibration} \textbf{333}, 1734 (2014).

\bibitem{shaw1983periodically}
S.~W.~Shaw and P.~J.~Holmes.
\newblock A periodically forced piecewise linear oscillator.
\newblock \emph{Journal of Sound and Vibration} \textbf{90}, 129 (1983).

\bibitem{andreaus2010numerical}
U.~Andreaus, L.~Placidi, and G.~Rega.
\newblock Numerical simulation of the soft contact dynamics of an impacting	bilinear oscillator.
\newblock \emph{Communications in Nonlinear Science and Numerical Simulation},
  \textbf{15} 2603 (2010).

\bibitem{zhang2011multi}
Y. Zhang, and K.D. Murphy.
\newblock Multi-modal analysis on the intermittent contact dynamics of atomic force microscope.
\newblock \emph{Journal of Sound and Vibration},
  \textbf{330} 5569 (2011).

\bibitem[Goldsmith(1964)]{goldsmith1964impact}
W.~Goldsmith.
\newblock \emph{Impact, the theory and physical behavior of colliding solids}.
\newblock (Edward Arnold Pub. Ltd., London, 1964).

\bibitem[Serweta et~al.(2014)Serweta, Okolewski, Blazejczyk-Okolewska,
  Czolczynski, and Kapitaniak]{serweta2014lyapunov}
W.~Serweta, A.~Okolewski, B.~Blazejczyk-Okolewska, K.~Czolczynski, and T.~Kapitaniak. 
\newblock Lyapunov exponents of impact oscillators with hertz's and newton's contact models. 
\newblock \emph{International Journal of Mechanical Sciences} \textbf{89}, 194, (2014).

\bibitem[Peterka(2003)]{peterka2003behaviour}
F.~Peterka. 
\newblock Behaviour of impact oscillator with soft and preloaded stop.
\newblock \emph{Chaos, Solitons \& Fractals} 18(1), 79 (2003).

\bibitem[Banerjee et~al.(1996)Banerjee, Bajaj, and Davies]{Banerjee1996}
B.~Banerjee, A.~K.~Bajaj, and P.~Davies. 
\newblock Resonant dynamics of an autoparametric system: A study using higher-order 	averaging. 
\newblock \emph{International Journal of Non-Linear Mechanics} \textbf{31}, 21 (1996).

\bibitem[Bernardo et~al.(2008)Bernardo, Budd, Champneys, and
  Kowalczyk]{bernardo2008piecewise}
M.~Bernardo, C.~Budd, A.~R.~Champneys, and P.~Kowalczyk.
\newblock \emph{Piecewise-smooth dynamical systems: theory and applications},
\newblock (Springer, 2008).

\bibitem[Ganguli and Banerjee(2005)]{ganguli2005dangerous}
A.~Ganguli and S.~Banerjee.
\newblock  Dangerous bifurcation at border collision: When does it occur?. \newblock \emph{Physical Review E} \textbf{71}, 057202 (2005).

\bibitem[Ma et~al.(2006)Ma, Agarwal, and Banerjee]{ma2006border}
Y.~Ma, M.~Agarwal, and S.~Banerjee. 
\newblock Border collision bifurcations in a soft impact system. 
\newblock \emph{Physics Letters A} \textbf{354}, 281 (2006).

\bibitem{blazejczyk2001antiphase}
B.~Blazejczyk-Okolewska, J.~Brindley, K.~Czolczynski, and T.~Kapitaniak.
\newblock Antiphase synchronization of chaos by noncontinuous coupling: two	impacting oscillators. \newblock \emph{Chaos, Solitons \& Fractals} \textbf{12},  1823 (2001).

\bibitem[Pikovsky et~al.(2001)Pikovsky, Rosenblum, and Kurths]{Pikovsky2001}
A.~Pikovsky, M.~Rosenblum, and J.~Kurths.
\newblock \emph{Synchronization. {A} Universal Concept in Nonlinear Sciences}.
\newblock (Cambridge University Press, 2001).

\bibitem[Strogatz et~al.(2005)Strogatz, Abrams, McRobie, Eckhardt, and
  Ott]{Strogatz2005}
S.~H.~Strogatz, D.~M.~Abrams, A.~McRobie, B.~Eckhardt, and E.~Ott.
\newblock Theoretical mechanics: Crowd synchrony on the millennium bridge.
\newblock \emph{Nature} \textbf{438}, 43 (2005).


\bibitem{Perlikowski2008a}
P.~Perlikowski, B.~Jagiello, A.~Stefanski, and T.~Kapitaniak.
\newblock Experimental observation of ragged synchronizability.
\newblock \emph{Physical Review E} \textbf{78}, 017203 (2008).

\bibitem{Blekhman1988}
I.I.~Blekhman,
\newblock Synchronization in Science and Technology,
\newblock ASME Press, New York, 1988.

\bibitem{Fujisaka1983}
H.~Fujisaka and T.~Yamada.
\newblock Stability theory of synchronized motion in coupled oscillator systems.
\newblock \emph{Progress Theoretical Physics} \textbf{69}, 32, (1983).

\bibitem{pikovsky1984}
M.~G.~Rosenblum, A.~S.~Pikovsky, and J.~Kurths.
\newblock Phase Synchronization of Chaotic Oscillators.
\newblock \emph{Physical Review Letters} \textbf{76}, 1804, (1996).

\bibitem{Kapitaniak20141}
M.~Kapitaniak, K.~Czolczynski, P.~Perlikowski, A.~Stefanski, and T.~Kapitaniak.
\newblock Synchronous states of slowly rotating pendula.
\newblock \emph{Physics Reports} \textbf{541}, 1 (2014).

\bibitem{pavlovskaia2010complex} 
E.~Pavlovskaia, J.~Ing, M.~Wiercigroch, and S.~Banerjee. 
\newblock Complex dynamics of bilinear oscillator close to grazing. 
\newblock \emph{International Journal of Bifurcation and Chaos} \textbf{20}, 3801 (2010).

\bibitem{menck2013basin}
P.~J.~Menck, J.~Heitzig, N.~Marwan, and J.~Kurths.
\newblock How basin stability complements the linear-stability paradigm.
\newblock \emph{Nature Physics} \textbf{9}, 89 (2013).

\bibitem{chudzik2011multistability}
A.~Chudzik, P.~Perlikowski, A.~Stefanski, and T.~Kapitaniak.
\newblock Multistability and rare attractors in van der Pol--Duffing oscillator.
\newblock \emph{International Journal of Bifurcation and Chaos} \textbf{21}, 1907 (2011).

\bibitem{brzeski2016}
P. Brzeski, M. Lazarek, T. Kapitaniak, J. Kurths, P. Perlikowski
\newblock Basin stability approach for quantifying responses of
multistable systems with parameters mismatch.
\newblock \emph{ArXiv e-prints} \textbf{1602.03751}, (2016).


\end{thebibliography}
\end{document}